\documentclass[aps,prl,twocolumn,showpacs,groupedaddress]{revtex4-1}
\usepackage{amsmath,amssymb,subfigure,wasysym}
\usepackage{color}
\usepackage{graphicx}
\usepackage{epstopdf}

\begin{document}

\title{Topological cascade laser for frequency comb generation in $\mathcal{PT}$-symmetric structures}

\author{Laura Pilozzi}\email{Corresponding author: laura.pilozzi@isc.cnr.it}
\affiliation{Institute for Complex Systems, National Research Council (ISC-CNR), Via dei Taurini 19, 00185 Rome, Italy}
\author{Claudio Conti}
\affiliation{Institute for Complex Systems, National Research Council (ISC-CNR), Via dei Taurini 19, 00185 Rome, Italy}
\affiliation{Department of Physics, University Sapienza, Piazzale Aldo Moro 5, 00185 Rome, Italy}

\begin{abstract}
The cascade of resonant topological structures with $\mathcal{PT}$-symmetry breaking is shown to emit laser light with a frequency-comb spectrum.
We consider optically active topological Aubry-Andr\'e-Harper lattices supporting edge-modes at regularly spaced frequencies.
When the amplified resonances in the $\mathcal{PT}$-broken regime match the edge modes of the topological gratings, we predict the emission of discrete laser lines. A proper design enables to engineer the spectral features for specific applications.
The robustness of the topological protection makes the system very well suited for a novel generation of compact frequency comb emitters for spectroscopy, metrology, and quantum information.
\end{abstract}


\maketitle
Applications  of frequency combs (FC)~\cite{Udem} are widespread and range from classical and quantum metrology to ultrafast spectroscopy. \cite{Diddams}
FC are coherent light sources with equally spaced spectral lines and are realized by highly nonlinear optical fibers, mode-locked lasers, and microring resonators. \cite{Jones,Herr12,Herr14,Kippenberg}
Compact and robust FC laser sources in different spectral ranges are subject to intense research and may potentially revolutionize photonics and quantum optics, including quantum cascade lasers and related terahertz and mid-infrared applications, as pollution detection and security. \cite{burghoff_terahertz_2014,rosch_octave-spanning_2015,hugi_mid-infrared_2012,combrie2016}
The key challenge for integrated FC sources is finding ideas to overcome limitations due to perturbation effects, as material disorder and
dispersion. From a fundamental point of view, classical and quantum properties of FC in complex structures are only marginally understood.\cite{Reimer1176}

Many authors \cite{Haldane,Raghu,Rechtsman,Khanikaev} recently applied topological concepts in optics,
and ``topological photonics'' is a rich and growing field.~\cite{Hasan, Lu} However, despite the potential robustness with respect to disorder
due to topological protection, FC laser emitters have not yet been considered because (i) resonant light-emitting topological systems have been only marginally studied, 
and (ii) edge-state laser sources, commonly studied in topological lasers, are intrinsically narrow-band, and one can argue about topological systems for broad-band operation.

In this Letter, we study cascaded topological resonances undergoing a $\mathcal{PT}$-symmetry breaking transition~\cite{Bender2,Makris} and demonstrate that it is possible to design a topologically protected FC active emitter. When the amplified resonances in the $\mathcal{PT}$-broken regime match the edge modes dispersion of the underlying topological lattice, we predict the emission of regularly spaced resonances. A designed $\mathcal{PT}$-symmetric topological-cascade emits frequency combs in target spectral ranges suitable, e.g., for dual-comb spectroscopy and related applications.

Proposals to use topological effects in the design of novel photonic lattices supporting many frequency channels for FC generation, with topological protection against disorder, have been recently performed in microring resonators~\cite{Ozawa,Yuan}. Topology allows to tune the competition with bulk extended modes and obtain single mode lasing on edge states of targeted wavelength. In fact the modulation of the structure produces gapless states whose localization at the edges induces a difference between the threshold gain of the targeted mode and its adjacent modes, giving excellent modal discrimination. Moreover by shaping the gain, parity-time symmetry further reduces the lasing threshold.

\begin{figure}[t]
\includegraphics[width=1.\columnwidth]{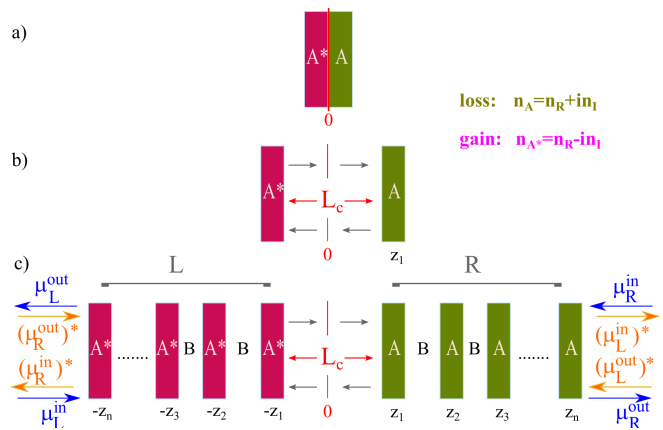}
\caption{(Color online)  One-dimensional $\mathcal{PT}$-symmetric topological structure after Eq.(\ref{eq:1}) and (\ref{eq:2}): (a) simplest case with $N_A=1$ and no central region; (b) as in (a) with central layer; (c) general case with input/output coefficients $\mu^{in,out}$ indicated. Blue (orange) arrows describes the configuration before (after) a $\mathcal{PT}$ transformation.\label{fig:lr2}}
\end{figure}
We here consider one-dimensional (1D) lattices~\cite{Lang,Ganeshan,Kraus,Posha,PoshaAr,PiCo1} with the Aubry-Andr$\acute{e}$-Harper (AAH) modulation~\cite{Aubry,Harper, Thou, Hofs, Guo1, Su}. We study resonant photonic structures~\cite{Posha} that have laser-light emitting topological edge states \cite{PiCo1}, which are symmetry protected with respect to structural perturbations at variance with surface states\cite{Dyakonov,kavokin,Shockley}. 
In our photonic AAH model, active layers $A$ are centered at positions $z_n = d_o\left({n + \eta \delta_n^H } \right)$ embedded in an homogeneous bulk material $B$. $\delta_n^H = \cos (2\pi\beta n + \phi )$ is the Harper modulation~\cite{Harper}, $d_o$ is the primary lattice period and $\eta$ is 
the modulation strength. For $\beta=p/q$ with $p>0$ and $q>0$ integers, the lattice displays two commensurate periods with $q$ resonant $A$ layers in the unit-cell. 
These resonant structures~\cite{Posha,PiCo1} are analogues of complex staggered potentials for electrons~\cite{Su}; a proper gain and loss distribution~\cite{Schomerus,Makris,Guo,Ruter,Longhi1} realizes synthetic optical media with parity-time ($\mathcal{PT}$) symmetry\cite{Zheng,Bender,Bender2}. This suggests 
adopting active components for lasers based on protected modes due to topological order\cite{Feng,Hodaei}.
\\
Letting $z$ the propagation direction, $\mathcal{PT}$-symmetry implies a constraint\cite{Makris} on the complex dielectric function $\varepsilon(z)=\varepsilon^*(-z)$.
This constraint is satisfied~\cite{El,Makris2} by alternating an even number of layers with $\varepsilon_n=\varepsilon_R+i(-1)^n\varepsilon_I$ with $(\varepsilon_R,\varepsilon_I)\in\mathbb{R}^+$ and $n=1,2,...$.  More complicated index distributions allow the balance between gain and loss around a symmetry point~\cite{Sean} and have a $\mathcal{PT}$-symmetric configuration with a rich phase diagram.
We consider the structure shown in Figure~\ref{fig:lr2} realized by joining a right region (R) with dielectric function
\begin{equation}
\begin{array}{l}
 \varepsilon_ > (z) = \sum\limits_{\ell  = 1}^{N_A} {\vartheta \left[ {z - (z_{\ell s} - L_A /2)} \right]} \vartheta \left[ {(z_{\ell s}  + L_A /2) - z} \right]\varepsilon _{\ell a}^ >   \\ 
  \quad + \sum\limits_{\ell  = 1}^{N_A} {\vartheta \left[ {z - (z_{\ell s}   + L_A /2)} \right]} \vartheta \left[ {(z_{(\ell  + 1) s}  - L_A /2) - z} \right]\varepsilon _{\ell b}^ >\text{,}   \\ 
 \end{array} \label{eq:1}
\end{equation}
with a left (L) region where:
\begin{equation}
\begin{array}{l}
 \varepsilon_ <  (z)=\\
  \sum\limits_{\ell  = -N_A}^{-1} {\vartheta \left[ {z - (z_{\ell (-s)}   - L_A /2)} \right]} \vartheta \left[ {(z_{\ell (-s)}   + L_A /2) - z} \right]\varepsilon _{\ell a}^ <   \\ 
  + \sum\limits_{\ell  = -N_A}^{-1}  {\vartheta \left[ {z - (z_{(\ell-1) (-s)}   + L_A /2)} \right]} \\
  \quad \quad \quad \quad \quad \quad \vartheta \left[ {(z_{\ell (-s)}  - L_A /2) - z} \right]\varepsilon _{\ell b}^ <.   \\ 
 \end{array} \label{eq:2}
\end{equation}
In Eqs.(\ref{eq:1},\ref{eq:2}), $\vartheta$ is the Heaviside step function, $\varepsilon _{\ell j}^ >   = \varepsilon _{Rj}  + i\gamma _{\ell j}$,  for $j=a,b$, and $\varepsilon _{\ell j}^ <   = \varepsilon _{Rj}  - i\gamma _{\ell j}$ so that, if the site in $z_n$ has a gain/loss coefficient $\gamma _{n j}$, the site in $-z_n$ has a loss/gain $-\gamma_{n j}$, thus realizing a $\mathcal{PT}$-symmetric structure. A central layer with length $L_c$ and real dielectric function $\varepsilon_c$ connects L and R regions. $N_A$ indicates the number of active A layers in L and R sections;
$\gamma_{n b}=0$ and $\gamma_{n a}\equiv \gamma_a$, $\forall~n$ hereafter. The shifted coordinates are $z_{\ell (\pm s)}=z_\ell \pm s$ with $s=(L_c+L_A)/2-z_1$.

We study the transition to broken $\mathcal{PT}-$symmetry by the eigenvectors and eigenvalues $s_\pm$ of the
scattering matrix $S$\cite{Chong}. $S$ maps incoming 
$\mu^{in}=\left(\mu_L^{in},\mu_R^{in}\right)$ to outgoing $\mu^{out}=\left(\mu_L^{out},\mu_R^{out}\right)$ coefficients by $\mu^{out}  = S \mu^{in}$. 
Blue and orange arrows in Fig.~\ref{fig:lr2} show the scattering configuration before (green) and after (orange) a $\mathcal{PT}$ transformation. In the $\mathcal{PT}$-symmetric configuration $\mu^{in}_L=(\mu^{out}_R)^*$ and $\mu^{in}_R=(\mu^{out}_L)^*$. For the $S$ eigenvectors such that $|\mu^{in}_{L,R}|=|\mu^{out}_{L,R}|$, 
one has $|\mu^{in}_L|=|\mu^{in}_R|$. Letting $\mu^{in}_L=1$ and $\mu^{in}_R= e^{i\phi }$ we have  \[
\left\{ \begin{array}{l}
 r_R e^{i\phi }  + t = 1 \\ 
 r_L  + te^{i\phi }  = e^{ - i\phi }  \\ 
 \end{array} \right.
\]
where $r_L$($r_R$) is the left (right) reflection amplitude and $t$ the transmission coefficient independent of the direction of incidence. 
Introducing the components $T_{ij}$ of the transfer matrix, one has $t(\omega)=1/T_{22}$, $r_L (\omega ) =  - T_{21}/T_{22}$, $r_R (\omega ) =T_{12}/T_{22}$ so that $\mathcal{PT}-$symmetry implies 
\begin{equation}
\xi (\omega ,\gamma_a ) \equiv \left| T_{21}+T_{12}\right| \le 2
\label{eq:eq2}
\end{equation}
with the equal sign corresponding to the symmetry breaking threshold.
\\
We analyze the structure in Fig.~\ref{fig:lr2}c exploiting the properties of the transfer and scattering matrices~\cite{Zyablovsky}: $T^*(\omega^*)=T^{-1}(\omega)$ and $(\mathcal{PT})S(\omega)(\mathcal{PT})=S^*(\omega^*)^{-1}$, and the symmetry relations between R and L sections. These allow to easily obtain transfer and scattering matrices of the whole $\mathcal{PT}$-symmetric structure from the knowledge of just the transfer matrix on the single period of the R section.
First we remark that, letting the phase shift $\phi  = \chi +\pi(1-2\beta)/2$, the structure is spatially inverted under the reversal $\chi\to-\chi$~\cite{Posha} so that the transfer matrix $T_L$ for the left-hand side is obtained by the corresponding right-hand side matrix $T_R$ by $T_L(\chi,\gamma)=T_R(-\chi,-\gamma)$.
The transfer matrix of the complete system is then $T_R(-\chi,-\gamma)\times~F^C\times~T_R(\chi,\gamma)$, where $F^C$ is the diagonal transfer matrix of the central layer with entries $F^C_{i,j}=\delta_{ij}\exp\left[(-1)^{i-1}i\omega \sqrt{\varepsilon_c} L_c/c \right]$ and $\delta_{i,j}$ the Kronecker delta.
Moreover the scattering matrix $S_R$ can be obtained by $T_R$ by simple components exchange\cite{Mostafazadeh}.
The left $S_L$ scattering matrix is then given by 
$S_L=\mathcal{PT}S_R=\sigma_x(S^*_R)^{-1}\sigma_x$ since:

- time-reversal $\mathcal{T}$ swaps incoming and outgoing states so that $\mathcal{T}S=(S^*)^{-1}$;

- parity $\mathcal{P}$ exchanges the left and right sides and gives $\mathcal{P}S=\sigma_xS\sigma_x$ where $\sigma_x$ is the Pauli matrix.

\begin{figure}[t]
\includegraphics[width=1.\columnwidth]{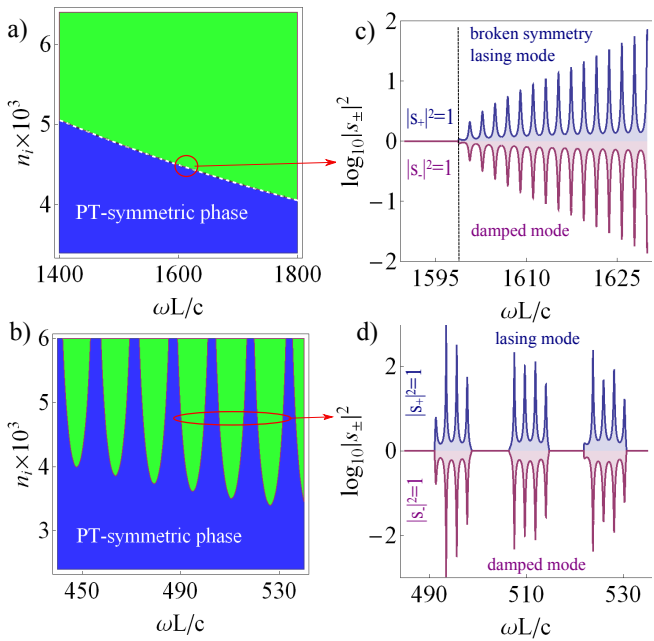}
\caption{(Color online)  a) Phase diagram for a pair of loss and gain media of total length L and $n_R=3$ (see Fig.\ref{fig:lr2}a); b) as in (a) for two layers with $n_R=3$ separated by a central layer of length $L_c=L/10$ and $n_c=2$ (total length $L$, see Fig.\ref{fig:lr2}b); c) semilog plot of S-matrix eigenvalue intensities ($log_{10}|s_\pm|^2$) versus normalized frequency for $n_i=4.5
 \times 10^{-3}$ for the two-layers; d) as in (c) for the three-layers and $n_i=4.3\times 10^{-3}$\label{fig:lg1}.}
\end{figure}

To gain insight into the physics of our $\mathcal{PT}$-symmetric topological structure, we first consider the simplest systems in Fig.~\ref{fig:lr2}. The two-layer structure in Fig.\ref{fig:lr2}a is made by a pair of loss (+) and gain (-) media with length $L/2$ and refractive index $n_{\pm}=n_R \pm in_i$; the structure in Fig.\ref{fig:lr2}b includes also a central layer. Figure~\ref{fig:lg1}a,b show the corresponding phase diagrams $\{\omega,n_i\}$. The dashed line in Figure~\ref{fig:lg1}a [equal sign in Eq.~(\ref{eq:eq2})] gives the symmetry breaking curve $n_i^c(\omega)$ in the absence of the central layer: the structure undergoes a phase transition when $\omega$ increases for a fixed gain/loss.
For the three-layer structure in Figure~\ref{fig:lg1}b, the phase diagram shows several broken-symmetry regions above a threshold value $n_i^{th}(\omega)$. Figure~\ref{fig:lg1}c,d show the eigenvalues $s_\pm$ of the scattering matrix. In the symmetric phase, one has two unimodular eigenvalues ($log_{10}|s_\pm|^2=0$). In the broken-symmetry regime, one finds two eigenvalues with reciprocal moduli ($log_{10}|s_+|^2=-log_{10}|s_-|^2$). For the two-layer structure, Fig.\ref{fig:lg1}c shows that above a critical frequency, one eigenvalue (blue curve) exhibits amplification. For the three-layer structure (Figure~\ref{fig:lg1}d) different regions with amplified modes are present.
Moreover, the eigenvalues $s_\pm$ define the threshold for lasing as a real frequency pole $\omega_o$ such that $ T_{22}(\omega_o)=0$. Due to $\mathcal{PT}$-symmetry $T^*(\omega^*)=T^{-1}(\omega)$, we have $T_{11}(\omega_o)=0$, i.e., the system also shows a real frequency zero. This implies that, at the laser threshold, a pole ($s_+ \to \infty$) and a zero ($s_ -   \to 0$) of the scattering matrix coincide~\cite{Longhi} and the system is also a coherent perfect absorber \cite{Chong2}.

\begin{figure}[b]
\includegraphics[width=1.\columnwidth]{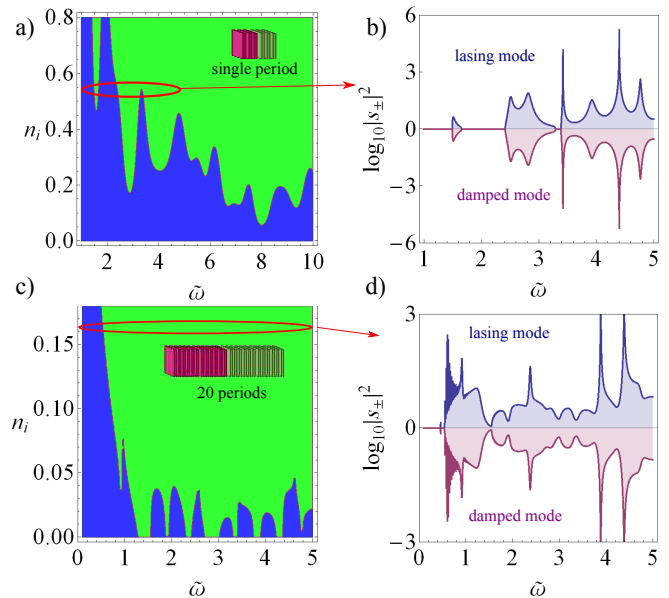}
\caption{(Color online) a) Phase diagram for $N_A=3$ with $\beta =1/3$ and $\phi =0.82\pi$; b) eigenvalues for $N_A=3$ when $n_i=0.53$; c) phase diagram for $N_A=60$ with $\beta =1/3$ and $\phi =0.82\pi$; d) eigenvalues for $N_A=60$ and $n_i=0.16$.\label{fig:fig3}}
\end{figure}

When increasing the number of real-index layers, internal resonances and oscillations in the non-hermiticity parameters produce a more complex phase diagram $\{\omega,n_i\}$~\cite{Chong}.
Letting $\tilde\omega=\omega d_o/c$, Fig.~\ref{fig:fig3}a shows $\{\tilde\omega,n_i\}$ for $N_A=3$ active layers in $z_n = d_o\left( {n + \eta \delta_n^H } \right)$ and Fig.~\ref{fig:fig3}b the eigenvalues for $n_i=0.53$. Fig.~\ref{fig:fig3}c shows the case $N_A=60$ and Fig.~\ref{fig:fig3}d the corresponding eigenvalues for $n_i=0.16$. Several amplified modes exist with a rich spectral distribution. ($L_c=d_o/10$).

To investigate the possibility of FC emission, we analyze the topological features of the complete structure in Fig.~\ref{fig:lr2}c. The two left (L) and right (R) sections exhibit localized left~($\ell$) and right~(r) modes with frequencies $w_{\ell(r)}^R=w_{r(\ell)}^L$ due to symmetry. This can be verified by the left-edge state frequencies $w_{\ell}^R$, which are the poles of the left reflection coefficient $R_L$~\cite{Posha}. When $\varepsilon_{Ra}\neq\varepsilon_{Rb}$, for $\beta=p/q$ with $p>0$ and $q>0$ integers, the spectral gaps of the un-modulated structure ($\eta=0$) at ${\tilde \omega} _B=\omega _B d_0/c =\pi /(\frac{{L_a}}{d_o}(\sqrt {\varepsilon _{Ra} } -\sqrt {\varepsilon _{Rb} })+\sqrt {\varepsilon _{Rb} } )$  split into $q$ gaps~\cite{Hofs}.
\begin{figure}[t]
\includegraphics[width=1.\columnwidth]{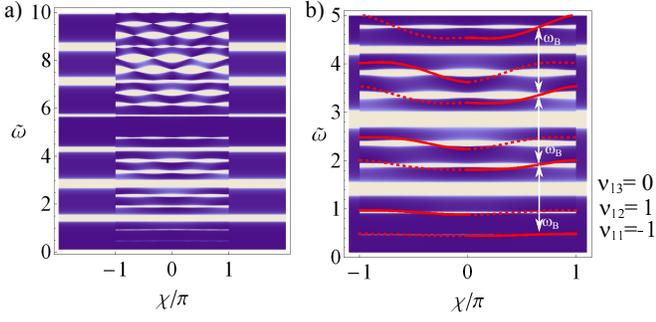}
\caption{(Color online) a) Reflectivity map for a structure with $\beta=1/3$. For $|\chi/\pi|>1$ one can identifies the bands of the un-modulated structure (bright regions). The range $|\chi /\pi| < 1$ shows the bands of the modulated structure; each stop band of the un-modulated structures ($|\chi/\pi|>1$) splits into $q=3$ bands; b) detail of reflectivity map in (a) showing left (continuous line) and right (dashed line) edge modes dispersion; for $q=3$ one has two edge-modes with $\nu_{ij}=\pm 1$. \label{fig:edge1}}
\end{figure}
Figure \ref{fig:edge1}a shows the absolute value of reflectivity for $\beta=1/3$ and reveals the allowed (dark) and forbidden (bright) bands for $\varepsilon _{Ra}= 9$, $\varepsilon _{Rb}= 4$ and $L_a/d_o=0.2$.
Each stop band, labeled by indexes $(i,j)$ with $i=1,2,..$ and $j=1,...,q$, is characterized by the topological invariant winding number
\begin{equation}
\nu_{ij}  = \frac{1}{{2\pi i}}\int\limits_{ - \pi }^\pi  {d\chi \frac{{\partial ln(r(\omega ,\chi ))}}{{\partial \chi }}}
\end{equation}
the extra phase in units of $2\pi$ of the reflection coefficient $r(\tilde\omega,\chi)$ when $\chi$ varies in the range ($-\pi,\pi$) with $\tilde \omega$ in the stop band~\cite{PoshaAr}.
For $\beta=1/3$ we have $\nu_{i1}=-1, \nu_{i2}=1, \nu_{i3}=0$, $\forall~i$. Due to bulk-boundary correspondence~\cite{Graf}, we have $|\nu_{ij}|=(1,1,0)$ edge modes in the corresponding gaps.
Figure~\ref{fig:edge1}b shows the dispersion of these edge-states localized at the left and right edges.
For a given $\chi$, the edge-states frequencies are equally spaced $\tilde \omega_n=\tilde \omega_o+n\tilde \omega_B$, as shown in details in Fig.~\ref{fig:edge3}a. These edge modes are amplified as they lay in the broken $\mathcal{PT}$-symmetry range (see the $log_{10}|s_+|^2$ curve in Fig.~\ref{fig:edge3}b).
This result is typical of the analyzed structures with the Harper modulation for any $\beta$, which can be varied to optimize the spectral emission.
\begin{figure}[t]
	\includegraphics[width=1.\columnwidth]{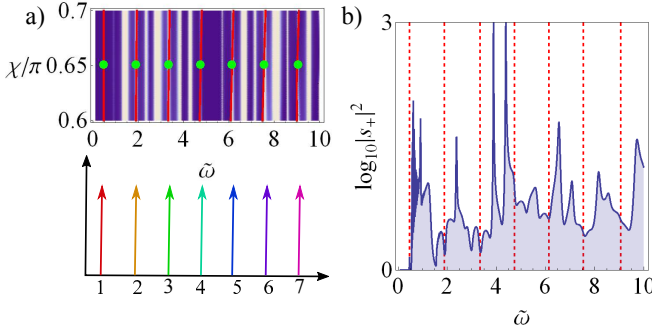}
	\caption{(Color online) a) Detail of the left edge modes dispersion in Fig.\ref{fig:edge1}b, bottom panel illustrates the FC; b) 
		$|s_+|$ eigenvalues for $n_i=0.16$, the red vertical lines mark the frequencies of the left localized edge modes revealing their amplification in the broken-symmetry regime.\label{fig:edge3}}
\end{figure}

\begin{figure}[t]
\includegraphics[width=1\columnwidth]{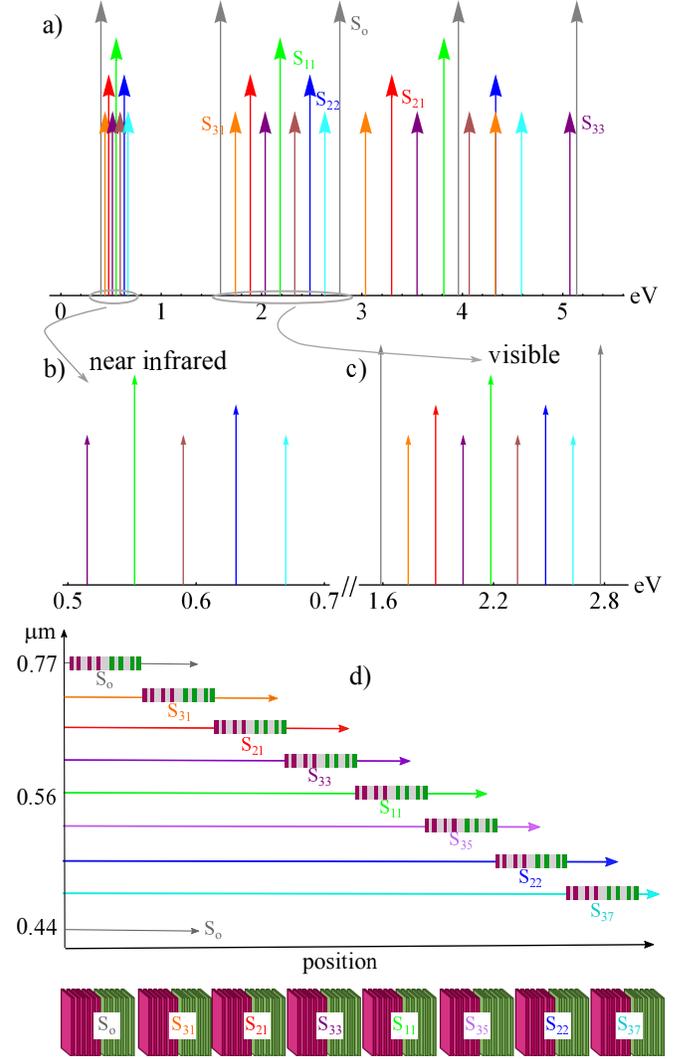}
\caption{(Color online) Spectral features of the FC emitted by cascaded $\mathcal{PT}$-symmetric topological systems: a) the structure $S_0$ emits the spectrum corresponding to the biggest gray lines; any additional structure $S_{ij}$ adds further FC lines and halves the spectral spacing in the visible range; b) details of the designed comb in the visible and near infrared; c) spatial distribution of the cascaded topological systems and corresponding lines in the visible. \label{fig:comb}}
\end{figure}

Our analysis shows that the considered system support a set of equally spaced topologically protected edge states that are amplified in the broken symmetry regime. A key result is that one can design the structure on the frequency range $(\omega_L,\omega_R)$ of interest. For a single structure with a given value $\chi$, the mode spacing, $\omega_B$ is much wider then the typical mode-spacing adopted in FC. However, one can realize any kind of FC by  {\it cascading} the single structure. By various $\mathcal{PT}$-topological structures $S_{ij}$ with different $d_{ij}$ values of the period $d_o$, one can obtain fractions of $\tilde \omega_B$ for the mode spacing. Figure~\ref{fig:comb} shows an example: a comb spectrum in the visible range $(\omega_L,\omega_R)=(1.59,2.778)~eV$. The design starts with a structure $S_o$ with parameters $d_o=237.1~nm$ and $L_a=47.4 nm$ such that the lowest $\tilde \omega_1=1.911~$ matches the $\omega_L=1.59~$eV value. This kind of structures can be realized by standard molecular beam epitaxy techniques as, e.g., for VCSEL laser employing AlGaAs as MQW for gain (as in ref.~\cite{arx}).
$S_0$ generates the fundamental comb shown as gray arrows in Fig.~\ref{fig:comb} with line spacing $\omega_B=1.19~$eV. A further cascaded structure $S_{11}$ with $d_{11}=\tilde \omega_1 c/(\omega_L+\frac{\omega_B}{2})$ adds the green lines and reduces the spacing in the range $(\omega_L,\omega_R)$ to $\omega_B/2$. 

This procedure can be iterated and each cascaded structure adds topologically protected lines to the emitted spectral comb. Indeed, recursively halving each range $[\omega_i,\omega_{i+1}]$ between two consecutive lines, 
$2^n$ equispaced modes in the range $(\omega_L,\omega_R)$ with spacing $ \omega_B/2^n$ can be obtained. This results in cascading $\sum\limits_{m  = 1}^n {2^{m-1}}$ $\mathcal{PT}$-topological structures to the initial $S_o$. For each structure $S_{ij}$, we have $d_{ij}=\tilde \omega_1 c/(\omega_L+j\frac{\omega_B}{2^i})$ for $i=(1,n)$ and odd $j=1,2^{n-1}$.
Figure~\ref{fig:comb}a shows the spectral features of the emission of seven cascaded $\mathcal{PT}$-symmetric topological systems. The overall structure sustains the emission of a FC in the visible range (Fig.~\ref{fig:comb}c) with line spacing 0.15~eV and, in addition, a FC in the near infrared range (Fig.~\ref{fig:comb}b) with line spacing 40~meV.

In conclusion, the interplay of gain and loss in a $\mathcal{PT}$-symmetric topological structure allows the amplification of multiple equally spaced frequencies when the resonances of lasing modes crosses the dispersion of edge-modes. A suitable design of the gain distribution of the topologically-protected edge-modes and the cascade of multiple lattices are the ingredients for realizing frequency-comb emitters with a prescribed spectral content. The inherent robustness of topological structures with respect to fabrication tolerances and the many degrees of freedom for design and optimization, including, e.g., spatial shaping of the gain profile, show the possibility of engineering specific applications, as dual-combs emitters or integrated optical clocks for metrology. Not only $\mathcal{PT}$-symmetry breaking in active topological systems reveals novel complex forms of light-matter interaction, but one can envisage a new generation of compact frequency-comb lasers and optically integrated sources of non-classical quantum states.

We acknowledge support from the Templeton foundation (grant number 58277) and from the project PRIN NEMO (reference 2015KEZNYM).

\end{document}